\documentstyle[12pt]{article}
\newcommand{\nc}{\newcommand}
\nc{\renc}{\renewcommand}

\nc{\etal}{\mbox{\it et al. }}
\nc{\ie}{{\it i.e.}}
\nc{\eg}{{\it e.g.}}

\renc{\thefootnote}{\arabic{footnote}}
\nc{\capt}[1]{{\bf Figure.} {\small\sl #1}}

\nc{\eqs}[2]{\mbox{Eqs.~(\ref{#1},\,\ref{#2})}}
\nc{\eq}[1]{\mbox{Eq.~(\ref{#1})}}

\nc{\figs}[2]{\mbox{Figs.~(\ref{#1},\,\ref{#2})}}
\nc{\fig}[1]{\mbox{Fig~.(\ref{#1})}}

\nc{\tag}[1]{\label{#1} \marginpar{{\footnotesize #1}}}
\nc{\mtag}[1]{\label{#1} \mbox{\marginpar{{\footnotesize #1}}}}
\renc{\baselinestretch}{1.2}
\newlength{\overeqskip}
\newlength{\undereqskip}
\nc{\be}[1]{\begin{equation} \mbox{$\label{#1}$}}
\nc{\bea}[1]{\begin{eqnarray} \mbox{$\label{#1}$}}
\nc{\Section}[2]{\section{#2}\label{#1}}
\nc{\Bibitem}[1]{\bibitem{#1}}
\nc{\Label}[1]{\label{#1}}

\nc{\eea}{\vspace{\undereqskip}\end{eqnarray}}
\nc{\ee}{\vspace{\undereqskip}\end{equation}}
\nc{\bdm}{\begin{displaymath}}
\nc{\edm}{\end{displaymath}}
\nc{\dpsty}{\displaystyle}
\nc{\bc}{\begin{center}}
\nc{\ec}{\end{center}}
\nc{\ba}{\begin{array}}
\nc{\ea}{\end{array}}
\nc{\bab}{\begin{abstract}}
\nc{\eab}{\end{abstract}}
\nc{\btab}{\begin{tabular}}
\nc{\etab}{\end{tabular}}
\nc{\bit}{\begin{itemize}}
\nc{\eit}{\end{itemize}}
\nc{\ben}{\begin{enumerate}}
\nc{\een}{\end{enumerate}}
\nc{\bfig}{\begin{figure}}
\nc{\efig}{\end{figure}}
\nc{\arreq}{&\!=\!&}
\nc{\arrmi}{&\!-\!&}
\nc{\arrpl}{&\!+\!&}
\nc{\arrap}{&\!\!\!\approx\!\!\!&}
\nc{\non}{\nonumber\\*}
\nc{\align}{\!\!\!\!\!\!\!\!&&}

\def\lsim{\; \raise0.3ex\hbox{$<$\kern-0.75em
      \raise-1.1ex\hbox{$\sim$}}\; }
\def\gsim{\; \raise0.3ex\hbox{$>$\kern-0.75em
      \raise-1.1ex\hbox{$\sim$}}\; }
\nc{\DOT}{\hspace{-0.08in}{\bf .}\hspace{0.1in}}
\nc{\Laada}{\hbox {$\sqcap$ \kern -1em $\sqcup$}}
\nc\loota{{\scriptstyle\sqcap\kern-0.55em\hbox{$\scriptstyle\sqcup$}}}
\nc\Loota{{\sqcap\kern-0.65em\hbox{$\sqcup$}}}
\nc\laada{\Loota}
\nc{\qed}{\hskip 3em \hbox{\BOX} \vskip 2ex}

\nc{\real}{{\rm I \! R}}
\nc{\Z}{{\sf Z \!\!\! Z}}
\nc{\complex}{{\rm C\!\!\! {\sf I}\,\,}}
\def\bigid{\leavevmode\hbox{\small1\kern-3.8pt\normalsize1}}
\def\id{\leavevmode\hbox{\small1\kern-3.3pt\normalsize1}}
\nc{\slask}{\!\!\!/}
\nc{\bis}{{\prime\prime}}
\nc{\pa}{\partial}
\nc{\na}{\nabla}
\nc{\ra}{\rangle}
\nc{\la}{\langle}
\nc{\goto}{\rightarrow}
\nc{\swap}{\leftrightarrow}

\nc{\EE}[1]{ \mbox{$\cdot10^{#1}$} }
\nc{\abs}[1]{\left|#1\right|}
\nc{\at}[2]{\left.#1\right|_{#2}}
\nc{\norm}[1]{\|#1\|}
\nc{\abscut}[2]{\Abs{#1}_{\scriptscriptstyle#2}}
\nc{\vek}[1]{\mbox{\bf #1}}
\nc{\veks}[1]{\mbox{\scriptsize\bf #1}}
\nc{\integral}[2]{\int\limits_{#1}^{#2}}
\nc{\inv}[1]{\frac{1}{#1}}
\nc{\dd}[2]{{{\partial #1}\over{\partial #2}}}
\nc{\ddd}[2]{{{{\partial}^2 #1}\over{\partial {#2}^2}}}
\nc{\dddd}[3]{{{{\partial}^2 #1}\over
	{\partial #2 \partial #3}}}
\nc{\dder}[2]{{{d #1}\over{d #2}}}
\nc{\ddder}[2]{{{d^2 #1}\over{d {#2}^2}}}
\nc{\dddder}[3]{{d^2 #1}\over
	{d #2 d #3}}
\nc{\dx}[1]{d\,^{#1}x}
\nc{\dy}[1]{d\,^{#1}y}
\nc{\dz}[1]{d\,^{#1}z}
\nc{\dl}[1]{\frac{d\,^{#1}l}{(2\pi)^{#1}}}
\nc{\dk}[1]{\frac{d\,^{#1}k}{(2\pi)^{#1}}}
\nc{\dq}[1]{\frac{d\,^{#1}q}{(2\pi)^{#1}}}

\nc{\cc}{\mbox{$c.c.$ }}
\nc{\hc}{\mbox{$h.c.$ }}
\nc{\cf}{cf.\ }
\nc{\erfc}{{\rm erfc}}
\nc{\Tr}{{\rm Tr\,}}
\nc{\tr}{{\rm tr\,}}
\nc{\pol}{{\rm pol}}
\nc{\sign}{{\rm sign}}
\nc{\bfT}{{\bf T }}

\nc{\cA}{{\cal A}}
\nc{\cB}{{\cal B}}
\nc{\cD}{{\cal D}}
\nc{\cE}{{\cal E}}
\nc{\cG}{{\cal G}}
\nc{\cH}{{\cal H}}
\nc{\cL}{{\cal L}}
\nc{\cO}{{\cal O}}
\nc{\cT}{{\cal T}}
\nc{\cN}{{\cal N}}
\nc{\rvac}[1]{|{\cal O}#1\rangle}
\nc{\lvac}[1]{\langle{\cal O}#1|}
\nc{\rvacb}[1]{|{\cal O}_\beta #1\rangle}
\nc{\lvacb}[1]{\langle{\cal O}_\beta #1 |}
\nc{\bb}{\bar{\beta}}
\nc{\bt}{\tilde{\beta}}
\nc{\ctH}{\tilde{\cal H}}
\nc{\chH}{\hat{\cal H}}
\nc{\1}{\aa}
\nc{\2}{\"{a}}
\nc{\3}{\"{o}}
\nc{\4}{\AA}
\nc{\5}{\"{A}}
\nc{\6}{\"{O}}
\nc{\al}{\alpha}
\nc{\g}{\gamma}
\nc{\Del}{\Delta}
\nc{\e}{\epsilon}
\nc{\eps}{\epsilon}
\nc{\lam}{\lambda}
\nc{\om}{\omega}
\nc{\Om}{\Omega}
\nc{\ve}{\varepsilon}
\nc{\mn}{{\mu\nu}}
\nc{\k}{\kappa}
\nc{\vp}{\varphi}

\nc{\advp}[3]{{\it  Adv.\ in\ Phys.\ }{{\bf #1} {(#2)} {#3}}}
\nc{\annp}[3]{{\it  Ann.\ Phys.\ (N.Y.)\ }{{\bf #1} {(#2)} {#3}}}
\nc{\apl}[3]{{\it  Appl. Phys. Lett. }{{\bf #1} {(#2)} {#3}}}
\nc{\apj}[3]{{\it  Ap.\ J.\ }{{\bf #1} {(#2)} {#3}}}
\nc{\apjl}[3]{{\it  Ap.\ J.\ Lett.\ }{{\bf #1} {(#2)} {#3}}}
\nc{\app}[3]{{\it Astropart.\ Phys.\ }{{\bf #1} {(#2)} {#3}}}
\nc{\cmp}[3]{{\it  Comm.\ Math.\ Phys.\ }{{ \bf #1} {(#2)} {#3}}}
\nc{\cqg}[3]{{\it  Class.\ Quant.\ Grav.\ }{{\bf #1} {(#2)} {#3}}}
\nc{\epl}[3]{{\it  Europhys.\ Lett.\ }{{\bf #1} {(#2)} {#3}}}
\nc{\ijmp}[3]{{\it Int.\ J.\ Mod.\ Phys.\ }{{\bf #1} {(#2)} {#3}}}
\nc{\ijtp}[3]{{\it Int.\ J.\ Theor.\ Phys.\ }{{\bf #1} {(#2)} {#3}}}
\nc{\jmp}[3]{{\it  J.\ Math.\ Phys.\ }{{ \bf #1} {(#2)} {#3}}}
\nc{\jpa}[3]{{\it  J.\ Phys.\ A\ }{{\bf #1} {(#2)} {#3}}}
\nc{\jpc}[3]{{\it  J.\ Phys.\ C\ }{{\bf #1} {(#2)} {#3}}}
\nc{\jap}[3]{{\it J.\ Appl.\ Phys.\ }{{\bf #1} {(#2)} {#3}}}
\nc{\jpsj}[3]{{\it J.\ Phys.\ Soc.\ Japan\ }{{\bf #1} {(#2)} {#3}}}
\nc{\lmp}[3]{{\it Lett.\ Math.\ Phys.\ }{{\bf #1} {(#2)} {#3}}}
\nc{\mpl}[3]{{\it  Mod.\ Phys.\ Lett.\ }{{\bf #1} {(#2)} {#3}}}
\nc{\ncim}[3]{{\it  Nuov.\ Cim.\ }{{\bf #1} {(#2)} {#3}}}
\nc{\np}[3]{{\it  Nucl.\ Phys.\ }{{\bf #1} {(#2)} {#3}}}
\nc{\pr}[3]{{\it Phys.\ Rev.\ }{{\bf #1} {(#2)} {#3}}}
\nc{\pra}[3]{{\it  Phys.\ Rev.\ A\ }{{\bf #1} {(#2)} {#3}}}
\nc{\prb}[3]{{\it  Phys.\ Rev.\ B\ }{{{\bf #1} {(#2)} {#3}}}}
\nc{\prc}[3]{{\it  Phys.\ Rev.\ C\ }{{\bf #1} {(#2)} {#3}}}
\nc{\prd}[3]{{\it  Phys.\ Rev.\ D\ }{{\bf #1} {(#2)} {#3}}}
\nc{\prl}[3]{{\it Phys\ Rev.\ Lett.\ }{{\bf #1} {(#2)} {#3}}}
\nc{\pl}[3]{{\it  Phys.\ Lett.\ }{{\bf #1} {(#2)} {#3}}}
\nc{\prep}[3]{{\it Phys\. Rep.\ }{{\bf #1} {(#2)} {#3}}}
\nc{\prsl}[3]{{\it Proc.\ R.\ Soc.\ London\ }{{\bf #1} {(#2)} {#3}}}
\nc{\ptp}[3]{{\it  Prog.\ Theor.\ Phys.\ }{{\bf #1} {(#2)} {#3}}}
\nc{\ptps}[3]{{\it  Prog\ Theor.\ Phys.\ suppl.\ }{{\bf #1} {(#2)} {#3}}}
\nc{\physa}[3]{{\it  Physica\ A\ }{{\bf #1} {(#2)} {#3}}}
\nc{\physb}[3]{{\it  Physica\ B\ }{{\bf #1} {(#2)} {#3}}}
\nc{\phys}[3]{{\it Physica\ }{{\bf #1} {(#2)} {#3}}}
\nc{\rmp}[3]{{\it  Rev.\ Mod.\ Phys.\ }{{\bf #1} {(#2)} {#3}}}
\nc{\rpp}[3]{{\it Rep.\ Prog.\ Phys.\ }{{\bf #1} {(#2)} {#3}}}
\nc{\sjnp}[3]{{\it Sov.\ J.\ Nucl.\ Phys.\ }{{\bf #1} {(#2)} {#3}}}
\nc{\spjetp}[3]{{\it Sov.\ Phys.\ JETP\ }{{\bf #1} {(#2)} {#3}}}
\nc{\yf}[3]{{\it Yad.\ Fiz.\ }{{\bf #1} {(#2)} {#3}}}
\nc{\zetp}[3]{{\it Zh.\ Eksp.\ Teor.\ Fiz.\  }{{\bf #1}  {(#2)} {#3}}}
\nc{\zp}[3]{{\it Z.\ Phys.\ }{{\bf #1} {(#2)} {#3}}}
\nc{\ibid}[3]{{\sl ibid.\ }{{\bf #1} {#2} {#3}}}
\nc{\rf}[1]{(\ref{#1})}
\nc{\nn}{\nonumber \\*}
\nc{\bfB}{\bf{B}}
\nc{\bfv}{\bf{v}}
\nc{\bfx}{\bf{x}}
\nc{\bfy}{\bf{y}}
\nc{\vx}{\vec{x}}
\nc{\vy}{\vec{y}}
\nc{\oB}{\overline{B}}
\nc{\oI}{\overline{I}}
\nc{\oR}{\overline{R}}
\nc{\rar}{\rightarrow}
\nc{\ti}{\times}
\nc{\slsh}{\hskip-5pt/}
\nc{\sm}{Standard~Model~}
\nc{\MP}{M_{\rm Pl}}
\nc{\tp}{t_{\rm Pl}}
\nc{\ave}{\bar{E}}

\renc{\min}{p_{\rm min}}
\renc{\max}{p_{\rm max}}
\nc{\pmin}{p_{\rm min}}
\nc{\pmax}{p_{\rm max}}
\nc{\fo}{f_0}
\nc{\foi}{f_{0,i}\,}
\nc{\fop}{f_0^P}
\nc{\fou}{f_0^U}
\def\sepand{\rule{14cm}{0pt}\and}
\nc{\eff}{{\rm eff}}
\nc{\MT}{M_{\rm T}}
\nc{\ML}{M_{\rm L}}
\nc{\kk}{\vek{k}}
\nc{\pp}{{\rm p}}
\nc{\cb}{critical bubble~}
\nc{\cbs}{critical bubbles~}
\nc{\scb}{subcritical bubble~}
\nc{\scbs}{subcritical bubbles~}

\begin{document}

{\title{{\hfill {{\small TURKU-FL-P17-95\\
	 \hfill hep-ph/9503483
        }}\vskip 1truecm}
{\bf Subcritical bubbles near the phase space domain wall}}


\author{
{\sc Jukka Sirkka$^{1}$}\\
 and \\
{\sc Iiro Vilja$^{2}$}\\
{\sl Department of Physics,
University of Turku} \\
{\sl FIN-20500 Turku, Finland} \\
\sepand
}
\maketitle}
\vspace{2cm}
\begin{abstract}
\noindent We study the subcritical bubble formation near the phase space
domain wall. We take into account that the phase of the scalar field can
vary using complex U(1) symmetric field and a phenomenological potential
with cubic term responsible to symmetry breaking. We show that the
presence of the domain wall induces subcritical bubbles so that their
formation rate near the wall is considerably larger than far of it.  The
allowed deviations of the phases of new bubbles are so large that they
prevent the system from induced nucleation.
\end{abstract}
\vfill
\footnoterule
{\small $^1$sirkka@utu.fi; $^2$vilja@utu.fi}
\thispagestyle{empty}
\newpage
\setcounter{page}{1}
The study of cosmological phase transitions have received considerable
attention in literature since the possibility of electroweak baryogenesis
was introduced \cite{CohenKN}. In that context the main effort has been
performed considering the critical configuration (bubble) formation which
are stable, expanding broken phase domains by their own. Less attention has
been paid to the vast subcritical bubble formation during the period of
the metastable state, because they have been supposed to have no effect
to the phase transition itself. Also it has become clear \cite{ studies1,
studies2, FodorH} that for physical Higgs mass, heavier that 60 GeV, the
electroweak phase transition is only weak. In the context of electroweak
 phase transition, subcritical bubbles were discussed first time by
\cite{GleiserKW} and it has been also shown \cite{janne} that phase
equilibrium can be reached at weak enough phase transitions. Kinetics of
subcritical bubbles has been studied in \cite{GelminiG} where phase
mixing above the critical temperature was investigated. More recently
numerical simulations of phase mixing in a 2+1 --dimensional model
\cite{Gleiser1} and in 3+1 --dimensional model \cite{BorrilG} have been
performed.

When the electroweak phase transition is weak enough and the formation
rate of broken phase configurations is large enough, there are large
number of subcritical bubbles present at and below the critical
temperature. It can happen that a collection of subcritical bubbles form
a region with a size of a critical bubble before they shrink away
\cite{EV}. Therefore the phase transition may be triggered by clustering
the broken phase regions. This mechanism of producing phase transition
is called induced nucleation.

In Ref. \cite{EV} it was demonstrated that in a simple model induced
nucleation is indeed possible for a large range of parameter values of
the scalar potential.  The growth of a (spherical) region of broken
phase was approximated to happen layer by layer by nucleating
subcritical bubbles.  As noted there, however, the probability that the
region of broken phase grows to the size of a \cb strongly depends on
the relative phases of the subcritical bubbles.  This effect was
accounted in \cite{EV} by introducing a parameter which describes the
probability that the phase of the nucleated \scb is correlated with the
phase of the pre-existing bubble. Its origin, however, was not further
analyzed there.

In the present paper we analyze the subcritical nucleation near phase
boundary. In particular, we concentrate to the question, how phases of
newly formed bubble and pre-existing broken phase domain correlate.  We
assume that the relative velocities between the bubbles are
negligible. In opposite case, the relative phase of colliding bubbles
would be washed out by thermal processes with a rate which is presumably
much larger than in the case of bubbles in rest, leading effectively to
the case where the phases of the bubbles are correlated.

One can also consider the effect of the distance $u$ of the \scb from
the pre-existing one. If $u$ is small, the newly formed bubble overlaps
significantly with the pre-existing one, causing the spatial phases to
be correlated. As $u$ increases, the spatial phase difference becomes
more probable.

We work with a $3+1$ --dimensional phenomenological model with a complex
scalar $\Phi$ having the $3$ --dimensional action
\be{aktio}
S_3[\Phi] = \int d^3x\, [|\na \Phi|^2 + V(\Phi)],
\ee
where the phenomenological potential is given by
\be{pot}
V[\Phi] = m(T)^2 |\Phi|^2 - \frac {2\sqrt 2 }3 \alpha T |\Phi|^3 +
\lambda |\Phi|^4.
\ee
Generally, the non-zero maximum $v_-$ and minimum $v_+$ of the potential
(below the critical temperature $T_c$) are given by
\be {ext}
v_\pm = {\alpha T\over 2\lambda}\left (1 \pm \sqrt{1 - \frac 89 \bar\lambda}
\right ) ,
\ee
where
\be{lviiva}
\bar\lambda = {9 \lambda m(T)^2\over 2\alpha^2 T^2}.
\ee
The parameter $\bar\lambda$ is less than unity at temperatures below $T_c$.

Generally there exist \scbs with various sizes and forms.
We assume that a typical subcritical bubble has a Gaussian,
spherically symmetric modulus
\be {bubble}
\Phi_b(x, y, z) = v_+(T) e^{- 2 \veks{x}^2/l^2}e^{i \chi(\veks{x})},
\ee
where the real function $\chi$ determines the phase of the bubble.
We use this configuration to represent all subcritical bubbles.
Subcriticality implies that the subcritical bubble diameter
$l \ll R_c$, the critical bubble radius.
Supposing that the pre-existing broken phase domain (bubble)
is large enough compared to the subcritical bubble its wall can be
approximated to be planar. Thus the background configuration is given by
\be {bg}
\Phi_{bg}(z) = v_+(T) e^{- 2 \theta(z)\, z^2/l^2},
\ee
where $\theta$ is the step function. The formation rate of subcritical
bubbles in the presence of $\Phi_{bg}$ at the distance $u$
is determined by the conditional probability and expressed as
\be {rate}
\Gamma_V[\chi] \simeq 
T^4\, e^{-\beta (S_3[\Phi_{bg}+\Phi_b^u]
-S_3[\Phi_{bg}])},
\ee
where the superscript $u$ denotes that the center of the bubble has been
moved to the point $\vek{x}=(0,0,u)$.

Now, the normalized probability density of phase configuration $\chi$ is
formally given by
\be {prob}
P[\chi] = {\cal N}^{-1}\Gamma'_V,
\ee
where
\be{rate'}
\Gamma'_V=e^{-\beta (S_3[\Phi_{bg}+\Phi_b^u]-S_3[\Phi_{bg}])}
\ee
and $ {\cal N} = \int {\cal D}\chi\, \Gamma'_V$. The evaluation of the
functional integral is, of course, a hopeless task.
In what follows we shall take some representative functions
$\chi$ to study statistical averages of some parameters using
$\Gamma'_V$ as a probability density.

To be specific, we write $\chi = \delta \chi'$ where $\delta$ is a
constant angle representing the size of the fluctuations and $\chi'$ is
one of the functions $\chi_1(z)=1/2(1+\tanh(z-u))$, $\chi_2(z)=
\theta(z)\theta(u-z)z/u+\theta(z-u)$ or $\chi_3(z)\equiv 1$.
Note that $\chi_1$ and $\chi_2$ have
asymptotic values 0 and 1 corresponding to the limits $z\goto -\infty$
and $z\goto \infty$, respectively. The function $\chi_3$ can be
thought to be the limiting case in such a class of functions.
Thus we can consider $\Gamma'_V$ as the probability density of the
random variables $u$ and $\delta$ with the normalization factor
\be {nor}
{\cal N'} =\int d u\, d\delta\, \Gamma'_V[\delta \chi']
\ee
Note, that the integrand is an even function of $\delta$.
The parameter $\delta$ can have any values in the cases of $\chi_1$ and
$\chi_2$ but by periodicity it is limited between $-\pi$ and $\pi$ in the
case of $\chi_3$.

The next step is to determine the maximal distance $u_{max}$.
In general $u$ is just a free parameter, but
e.g. in the case of induced nucleation the new bubbles can not be formed
arbitrary distant from the wall. One has to require that the new bubbles
join to the pre-existing one forming one connected domain. Let us for the
moment assume
that the all bubbles, both the pre-existing one and the newly formed ones,
have same phase; an assumption which certainly gives a maximal distance.
Suppose also that the pre-existing bubble has been formed layer by layer
from subcritical bubbles. By packing subcritical bubbles so that the
minimum in the middle of the bubbles the field value is larger that $v_-$,
which is required
that the field at that point would roll down to the broken minimum, not
towards unbroken one, one finds maximal possible $u$
\be {umax}
{u_{max}\over l_c} = \frac 23 \left [ 2\ln \left ({4v_+\over v_-}\right )
\right ]^{1/2},
\ee
corresponding to the tightest (lattice) packing of spheres in 3 --dimensions,
face centered cubic lattice \cite{ConwayS}. However, at $T_c$, where
$v_+/v_- = 2$ simple cubic lattice is equally good leading to same
$u_{max} = 1.36 l_c$ but for $v_+/v_- > 2$ it results a smaller one.
The mean phase fluctuation size $\bar \delta$ can now be given by
\be {deltabar}
\bar\delta^2 = \int d\delta\integral{0}{u_{max}}du\, \delta^2
P(u,\delta),
\ee
and similarly for $\bar u$.

We are mostly interested in weak phase transitions where
the actual transition temperature $T_f$ is close to the critical
temperature $T_c$. Thus we, as an approximation, perform the analysis at $T_c$
except in some formulas where the tiny difference between $T_c$ and $T_f$ is
crucial. At small supercooling limit $T_f$ is determined by \cite{janne}
\be {transT}
S^b_3/T = {\alpha\over\lambda^{3/2}}{2^{9/2}\pi\bar\lambda^{3/2}\over 3^5
(\bar\lambda - 1)^2} \simeq 150
\ee
leading to $\bar\lambda = 1 - 0.0442 \alpha^{1/2}/\lambda^{3/4}$. Small
supercooling limit is valid if $1 - \bar\lambda \ll 1$, i.e.
$\alpha/\lambda^{3/2}\ll 500$. The minimum at $T_f$ is given by
\be {vminatTf}
-\epsilon \equiv V(v_+(T_f), T_f)
\simeq -0.00218 {\alpha^{9/2}\over \lambda^{15/4}}T_c^4.
\ee

At the critical temperature the non-zero minimum of the potential reads
\be {vac}
v_+(T_c) = {\sqrt 2\over 3}{\alpha\over\lambda}T_c
\ee
and the mass is given by
\be {mass}
m(T_c)^2 = {2\over 9}{\alpha^2\over\lambda}T_c^2.
\ee
The correlation length in symmetric phase is
\be {corrl}
l_c = {1\over m(T)}
\ee
and, although all sizes of subcritical bubbles exists in the symmetric phase,
we use a representative subcritical bubble\cite{janne, EV} with $l = l_c$.

The effect of approximation $T_f\simeq T_c$ and use of $l=l_c$ is that
the quantity $\beta S_3$ depends on the potential parameters only through
the combination $\gamma\equiv\alpha/\lambda^{3/2}$.
The numerical evaluation of the action shows, provided that
$\gamma$ is not too close to zero i.e. the phase transition is not too
weak, that the main contribution
to the averages $\bar\delta^2$ and $\bar u$ comes from the region where
$\delta$ is close to zero
and $u$ is close to $u_{max}$ in the cases of functions $\chi_1$ and
$\chi_2$.
But in the case of $\chi_3$ the averages receive also a
significant contribution from the region where $\delta$ is close to $\pi$
and $u$ is close to zero, as shown in Fig. 1. This is to be interpreted that
in the case of the
constant phase there are in average two kinds of subcritical bubbles:
those having
almost opposite phase as the pre-existing bubble and lying near the wall
and those having almost the same phase as the pre-existing bubble and
lying as far as possible from the wall. For $\chi_1$ and $\chi_2$ the averages
are plotted in Fig. 2.

By comparing the average formation rate $\overline {\Gamma}'_V$ of subcritical
bubbles near the domain wall to the rate $\Gamma_s$ of a single bubble not
influenced by the wall given by \cite{EV} $\Gamma_s \propto e^{- 2.06 \gamma
}$ we find that for values $\gamma > .5$ the rate is significantly larger near
the domain wall, Fig. 3. Thus the domain wall is shielded by the subcritical
bubbles nearby it. The shielding is even stronger for larger values of
$\gamma$. This would enhance the induced nucleation by making subcritical
bubble nucleation easier. However, the relative phases of the bubbles do not
correlate any more and therefore we have to estimate the influence of the
phase differences.

The influence of phase deviations of the size $\bar\delta$ to the bubble
dynamics can be most easily analyzed within the context of thin wall
approximation valid in weak phase transition case. Thin wall approximation
requires that the maximum of the potential $V_{max} \gg \epsilon$, which
implies $\alpha/\lambda^{3/2} \ll 10$ in accordance with small supercooling
limit. The critical bubble
radius $R$ is obtained by extremizing the bounce action
\be {bounce}
B = - \frac 43 \pi \epsilon_{eff} R^3 + 4 \pi \sigma R^3,
\ee
where the surface tension in thin wall approximation is given by
\be {surface}
\sigma = {2\sqrt 2 \alpha^3\over 91 \lambda^{5/2}}T_c^3.
\ee
Usually $\epsilon_{eff}$ is simply given by \eq{vminatTf} but now it is
affected by phase fluctuations in the broken phase, so that
\be {epseff}
\epsilon_{eff} = \epsilon - \Delta\epsilon,
\ee
where
\be {deltaeps}
\frac 43 \pi R^3\Delta\epsilon = \langle \!\! \integral{\veks{x}
\in B(0, R)} {}\!\! d^3x\, |\na \Phi_R(\vek{x})|^2\rangle.
\ee
$\Phi_R$ is now a random configuration having modulus $v_+$ and
a varying phase between values $-\bar\delta$ and $\bar\delta$.

If we suppose that the typical size of the spatial extension of the
phase fluctuations is $\bar u$, as can be done in the cases of $\chi_1$ and
$\chi_2$, we are able to
calculate $\Delta\epsilon$. When the maximum values of
phase fluctuations are situated at the points $\vek{x} =
(n_x\bar u, n_y \bar u, n_z\bar u),\ n_x,\ n_y,\ n_z \in \Z$ the calculation
can be reduced to
a unit lattice where $n_x,\ n_y,\ n_z = 0,\ 1$. Taking $\Phi_R =
e^{i\bar\delta \chi(\veks x)}$ the expectation value
$\Delta\epsilon$ can be now expressed in terms of the unit lattice:
\be {de}
\frac 43 \pi R^3 \Delta\epsilon = \frac 43 \pi v_+^2 \bar\delta^2
\left (\frac R{\bar u} \right)^3 \langle \!\! \integral{\veks{x}
\in [0,\ \bar u]^3} {} \!\! d^3x\, (\na \chi)^2\rangle.
\ee
At each corner we associate equally distributed independent random variables
$f_{n_x,n_y,n_z}$ with values  $\pm 1$ and correlations
$\langle f_{n,n,n}^N\rangle = {1 +(-1)^N\over 2} $.
Moreover, we choose the phase
configuration in the unit bravais lattice to be the simplest possible
polynomial, i.e. the Lagrangian interpolation polynomial
$$
\chi(\vek x ) = - \!\! \sum_{n_x, n_y, n_z = 0,\ 1}\!\!\!f_{n_x, n_y, n_z}
\!\!\!\!\!\!\prod_{\mbox{\it\scriptsize
\begin{tabular}{c} {$k_x, k_y, k_z = 0,\ 1$}\\
                   {$k_x\neq n_x$}\\
                   {$k_y\neq n_y$}\\
                   {$k_z\neq n_z$}
\end{tabular}}}\!\!\!\!\!\!
(-1)^{k_x + k_y + k_z}(k_x  - \frac x{\bar u} )(k_y -\frac y{\bar u} )
(k_z -\frac z{\bar u}).
$$
An easy calculation yields now
\be {result}
\Delta\epsilon = {8 v_+^2 \over 3 \bar u^2}\, \bar\delta^2\, .
\ee
Because from \eq{bounce} one obtains the critical bubble radius
\be {Rc}
R_c = {3\sigma\over \epsilon_{eff}}
\ee
the critical bubble radius tends to infinity if $\epsilon_{eff}\goto 0$.
Hence a general requirement for existence of the critical bubble can be
stated to be $\epsilon_{eff} > 0$, i.e.
\be {final}
\bar\delta < 0.129 {\bar u\over l_c}\left ({\alpha\over\lambda^{3/2}}
\right )^{1/4}.
\ee
Thus \eq{final} expresses an overall upper bound for mean phase fluctuations
inside a critical bubble during the cosmological electroweak phase transition
in terms of phenomenological parameters and fluctuation correlation
length. Comparing Eq. (\ref{final}) and the Fig. 2 one can conclude that it is
not possible for any size of critical bubble to survive from the effects of
phase fluctuations. One should have a mechanism which would damp more
efficiently the variations of $\delta$.

In the case of $\chi_3$ the energy change has to be calculated differently.
Because the small, opposite phase $\delta = \pi$ bubbles can be approximated
to be isolated in the $\delta = 0$ phase, the energy change is approximated by
$\delta \epsilon = {E_2\over v_1 + v_2},$ where $E_2$ is the energy of single
wrong-phase bubble and $v_1$ and $v_2$ are the volumes of $\delta = 0$ and
 $\delta = \pi$ bubbles, correspondingly. A calculation yields an lower bound
for $\delta \epsilon > 0.0595\gamma$. Requiring $\epsilon_{eff} > 0$ we obtain
$\alpha\lambda^{1/2} > 0.55$ or, because $\gamma \ll 10$, $\lambda \gg 0.23$
which can hardly be true. Hence also $\chi_3$ case confirms the conclusion
that the phase deviations are too large for induced nucleation to be possible.

In the present paper we have analyzed the subcritical bubble nucleation near a
domain wall. The wall itself induces nucleation nearby it so that the
subcritical bubble formation rate is larger close the wall than far of it. It
also appears that deviations of the phase of the field can be remarkably large
and prohibits the system from induced nucleation. Note, that if the size of
typical subcritical bubble is smaller that the correlation length as suggested
lately \cite{SMY}, the bound (\ref{final}) would be even stricter. Also those
subcritical bubbles having very small amplitude compared to $v_+$ may have
effect to the large amplitude configurations. In realistic theories, like the
Standard model where the symmetry group is larger, the actual numbers for
$\bar \delta$ and $\bar u$ would surely alter but, because they have more
degrees of freedom, the qualitative results are likely to hold.
 \newpage
\noindent {\Large{\bf Figure captions}} \vskip .5truecm

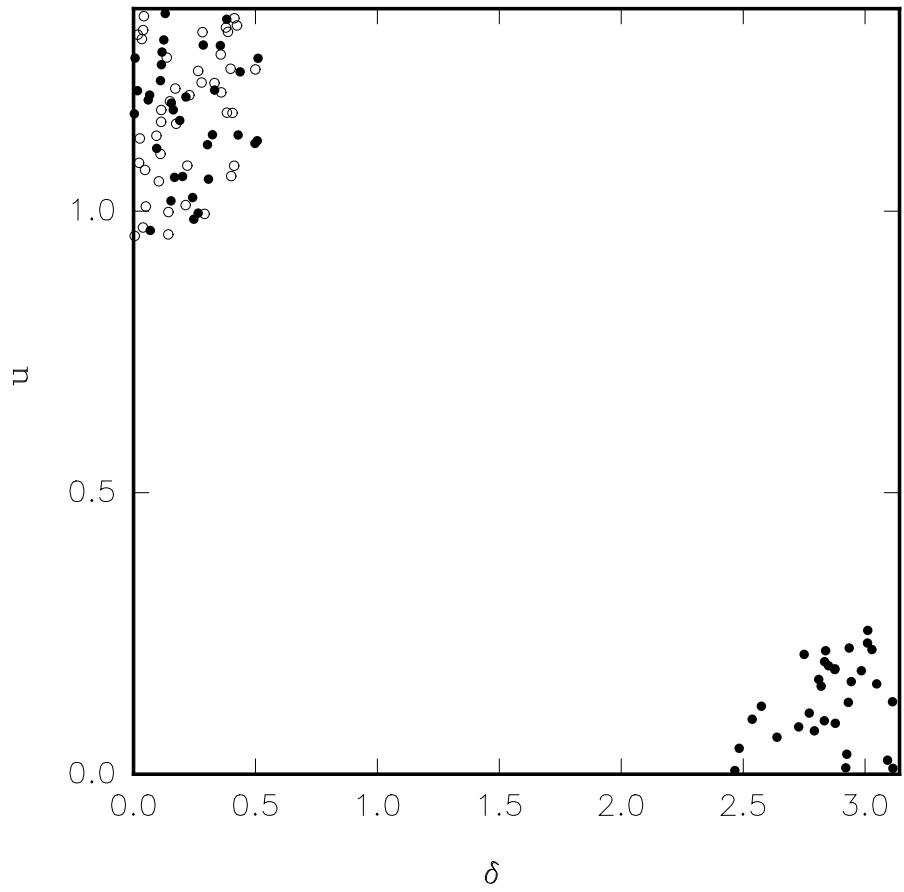
\noindent Figure 1. The fraction of 500 randomly generated points in the
$\delta$-$u$ plane which pass the limit $\beta S_3<1$ with $\gamma=1$. Open
circles correspond to the case of $\chi_1$ and filled circles
to the case of $\chi_3$.
\vskip .5truecm

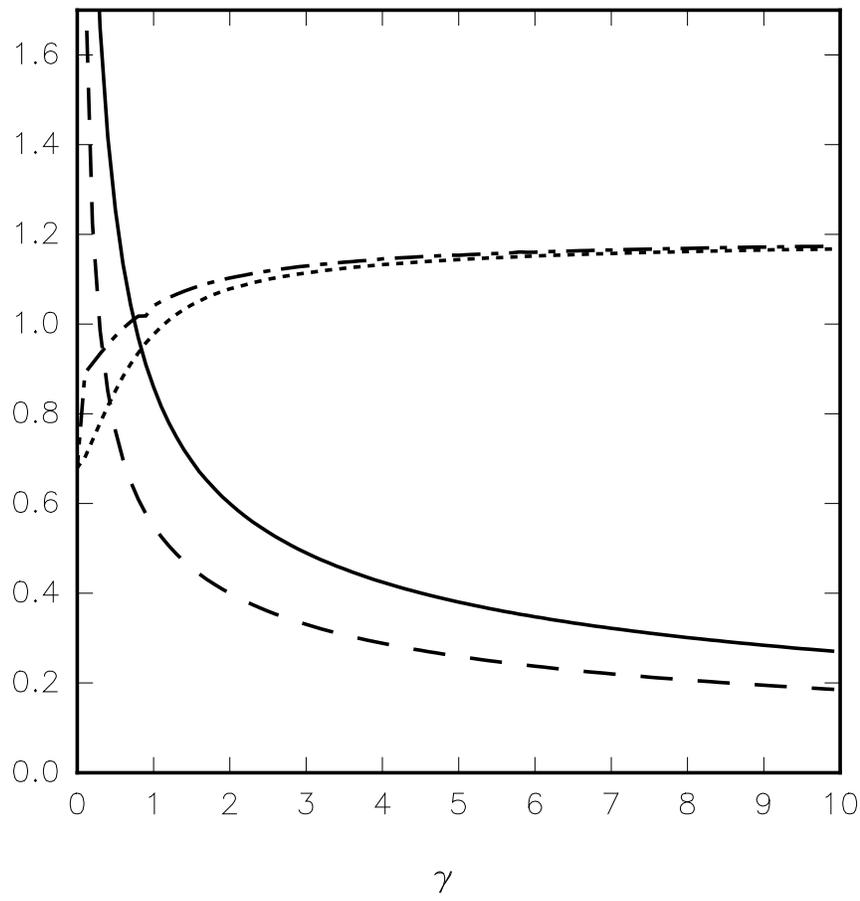
\noindent Figure 2. The averages $\bar{\delta}[\chi_1]$ (solid line),
$\bar{\delta}[\chi_2]$ (dashed line), $\bar{u}[\chi_1]$ (dotted line) and
$\bar{u}[\chi_2]$ (dash-dotted line) as a function
of $\gamma$. For $\gamma=0$ the average $\bar\delta$ diverges as a function
of the upper limit of the $\delta$-integration.
The upper limit used in the plot is $\delta_{max}=10^2$. Studying the
dependence of the upper limit of the integration numerically reveals that
it is wery weak for $\gamma\gsim 0.1$, implying that only those values
of $\bar\delta$ with $\gamma$ larger than that limit are reliable.
Note that $\gamma=0$ implies that the scalar potential has only the symmetric
phase minimum $|\Phi|=0$.
\vskip .5truecm

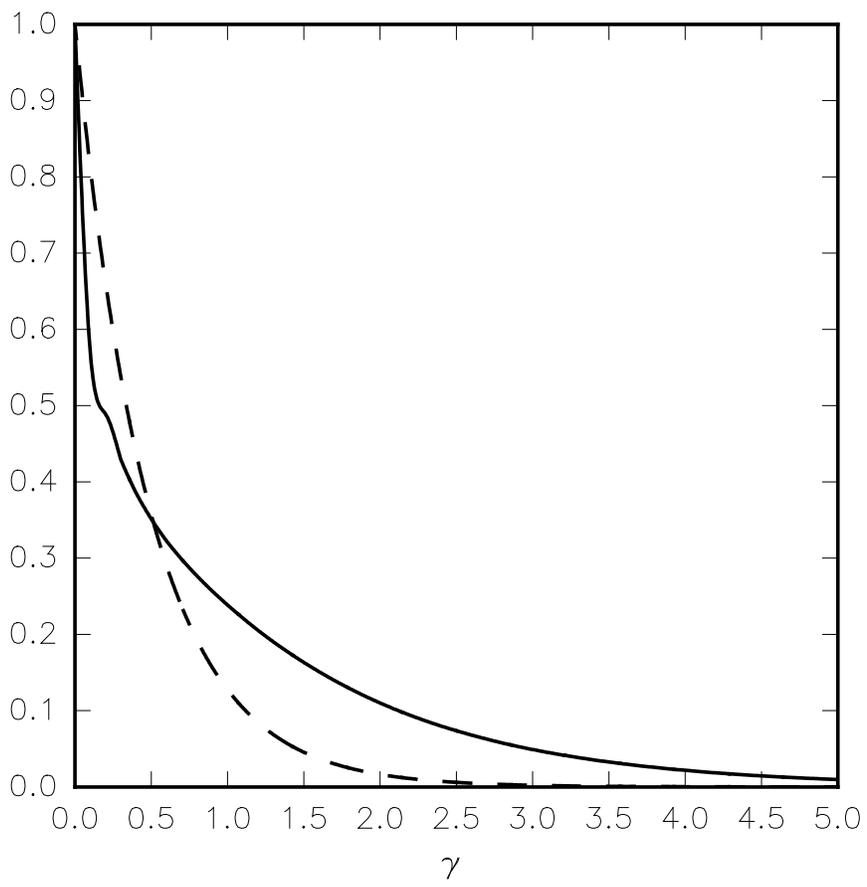
\noindent Figure 3. The average formation rate $\overline{\Gamma}'_V$ of \scbs
near the domain wall (solid line) and the rate $\Gamma_s=\exp(-2.06\gamma)$
of a single bubble not influenced by the wall (dashed line).

\end{document}